\documentclass{emulateapj}

\newcommand{\diff}[2]{\frac{\partial #1}{\partial #2}}

\newcommand*{\eps}{\epsilon}

\newcommand*{\etal}{et al.\ }

\shorttitle{Spectral Hardening in Large Solar Flares}

\begin{document}
%

\title{Spectral Hardening in Large Solar Flares}

\author{P. C. Grigis\altaffilmark{1,2}}
\affil{pgrigis@cfa.harvard.edu}
\and 
\author{A. O. Benz\altaffilmark{1}}
\affil{benz@astro.phys.ethz.ch}
\altaffiltext{1}{ETH Zurich, 8092 Zurich, Switzerland}
\altaffiltext{2}{Harvard-Smithsonian Center for Astrophysics, Cambridge MA 02138, USA}

\begin{abstract}

  Observations by the Ramaty High Energy Solar Spectroscopic Imager (RHESSI) are
  used to quantitatively study the hard X-ray evolution in 5 large solar flares
  selected for spectral hardening in the course of the event. The X-ray
  bremsstrahlung emission from non-thermal electrons is characterized by two
  spectroscopically distinct phases: impulsive and gradual. The impulsive phase
  usually consists of several emission spikes following a soft-hard-soft
  spectral pattern, whereas the gradual stage manifests itself as spectral
  hardening while the flux slowly decreases.

  Both the soft-hard-soft (impulsive) phase and the hardening (gradual) phase
  are well described by piecewise linear dependence of the photon spectral index
  on the logarithm of the hard X-ray flux. The different linear parts of this
  relation correspond to different rise and decay phases of emission spikes.
  The temporal evolution of the spectra is compared with the configuration and
  motion of the hard X-ray sources in RHESSI images.

  These observations reveal that the two stages of electron acceleration causing
  these two different behaviors are closely related in space and time. The
  transition between the impulsive and gradual phase is found to be smooth and
  progressive rather than abrupt. This suggests that they arise because of a
  slow change in a common accelerator rather than being caused by two
  independent and distinct acceleration processes. We propose that the hardening
  during the decay phase is caused by continuing particle acceleration with
  longer trapping in the accelerator before escape.

\end{abstract}


%
\section{Introduction}
%

Large solar flares are very bright hard X-ray sources. The emission originates
from energetic electrons with energies mainly in the 10s and 100s of keV,
believed to be accelerated in the corona. These electrons have a short lifetime
in regions dense enough to generate substantial hard X-ray emission, and
therefore react quickly to changes in the acceleration, transport and emission
processes. While it may be hard do disentangle the contributions of the various
effects in different flares, the investigation of the temporal evolution
of the hard X-ray spectra in single flares is a valuable tool to study these
processes. Flare models and theories should be able to account for the behavior
of the observed hard X-ray spectra as they change during an event.

Observations of the spectral hard X-ray evolution have revealed two main
trends: a soft-hard-soft (SHS) spectral evolution of emission peaks, and a
progressive hardening during whole events (SHH, soft-hard-harder).

The SHS behavior of emission spikes was discovered by \cite{parks69}, and since
then has been reported by many others. Recently, \cite{grigis04} surveyed
quantitatively the spectral evolution of emission spikes during M class events,
finding that nearly all rise and decay phases of the peaks show the SHS
behavior. The excursions in both photon flux measured at a fixed energy and
spectral hardness can be very different from peak to peak. However, they show
consistently a characteristic property: the spectral power-law index is a linear
function of the logarithm of the flux \citep{grigis05}. The SHS pattern has also
been observed in looptop sources \citep{battaglia06}. Thus it is likely to be a
characteristic signature of acceleration rather than a propagation
process. Detailed modeling of transit-time damping acceleration of electrons can
reproduce the SHS behavior if the effects of particle trapping in the
acceleration region and escape are taken in account \citep{grigis06}.

The SHH behavior was first observed by \citet{frost71}, who noted that the
spectral index in the late phase of a flare stayed constant at a harder value
than measured during the first (impulsive) SHS peak. Further events were studied
by \citet{cliver86} and \citet{kiplinger95} using data from the Hard X-Ray Burst
Spectrometer (HXRBS) on SMM. The distinctive feature of the SHH evolution is the
absence of softening as the flux decreases.

Kiplinger found two different subtypes of behavior:
\begin{itemize}
\item hardening during a particular peak
\item hardening during the decay of the whole event
\end{itemize}
In the first subtype, substantial hardening occurs during a short period, but
after an emission peak the flux may soften again. Events of the second subtype
(corresponding to the classic flare with a gradual phase) typically have some
SHS peaks at the beginning but progressively harden afterwards. Despite the name
SHH, a general hardening may start already before the largest peak. Thus SHH is
not limited to the decay phase. We note also that the two classes of spectral
evolution are not clearly separated: most SHH events show impulsive SHS peaks in
the beginning. The interest in SHH flares rose after Kiplinger's report of a
high association rate between SHH and the occurrence of interplanetary energetic
proton events. More recently, \citet{2008ApJ...673.1169S} confirmed the
association between solar energetic particles and hardenings during the January
2005 solar storm events. The link between the two phenomena, however, is 
not the subject of this work.

The two different kinds of spectral evolution (SHS and SHH) seem to support the
view that there are two different stages in flares: an impulsive phase at the
beginning followed by a gradual component, corresponding to different
acceleration mechanisms. This scenario was first proposed to explain radio
observations \citep{wild63}. A first (impulsive) phase was suggested to
accelerate electrons producing gyrosynchroton emission, and the second (gradual)
phase was linked to traveling shocks (type II radio bursts) accelerating further
the electrons and also ions. This idea was then used to interpret hard X-ray
observations \citep{frost71, bai1979}. Later, the shocks were associated with Coronal Mass
Ejections (CMEs). Occulted flares seemed to confirm this scenario
\citep{hudson82}. However, it cannot explain the position of the dominant hard
X-ray source seen during the gradual phase: imaging observations by Hinotori
\citep{ohki83} showed that the hard X-ray emission comes from too low in the
solar corona to justify the connection with type II radio
bursts. \citet{kahler84} and  \citet{bai1986} argued that the impulsive phase is followed by two
independent acceleration processes. The first happening in the post-flare loop
arcade is responsible for the late-phase hard X-ray emitting electrons, and the
second higher up in the corona, shock driven, accelerates interplanetary
electrons and ions. This was later corroborated by \citet{cliver86} using SMM
observations. 

Stochastic acceleration reproduces the observed SHS behavior but 
cannot at the same time describe hardening when the flux decays. If both SHS and
SHH phases of electron acceleration happen in the same event, why does the
spectral behavior reverse? Are different acceleration mechanisms at work, or is
there a further parameter in the same process that changes in the course of the
flare? To find an observational answer to this question, simultaneous imaging
and spectral observations are analyzed that have become available for the first
time by the Reuven Ramaty High Energy Solar Spectrometric Imager
\citep[RHESSI;][]{lin02}. Its data characterize the spectral evolution of the
non-thermal hard X-ray flux in unprecedented detail. We also compare the
spectral evolution with the geometrical flare configuration, as well as the
motions of the coronal X-ray source and the non-thermal footpoint sources.

%
%
 
\begin{table}
\caption{Chronological list of the selected events.}
\label{tab:evlist2}
\centering
\begin{tabular}{r c c l}
\hline\hline
\multicolumn{1}{c}{Date}
& \multicolumn{2}{c}{\hspace{0.3cm}GOES peak\hspace{0.6cm} } & Panel in \\
      & time & flux  & Figs. \ref{fig:lightcurves} and \ref{fig:glogf}\\
\hline
  7-NOV-2004 & 16:05 & X2.0 & F\\ 
 10-NOV-2004 & 02:13 & X2.6 & A\\ 
 17-JAN-2005 & 09:52 & X3.9 & B\\ 

 19-JAN-2005 & 08:23 & X1.4 & C \& D\\ 
 20-JAN-2005 & 07:01 & X7.1 & E\\ 
\end{tabular}
\end{table}
 
%
%

%
\section{Method}
\label{methods}

%
%
 
\begin{figure}
\plotone{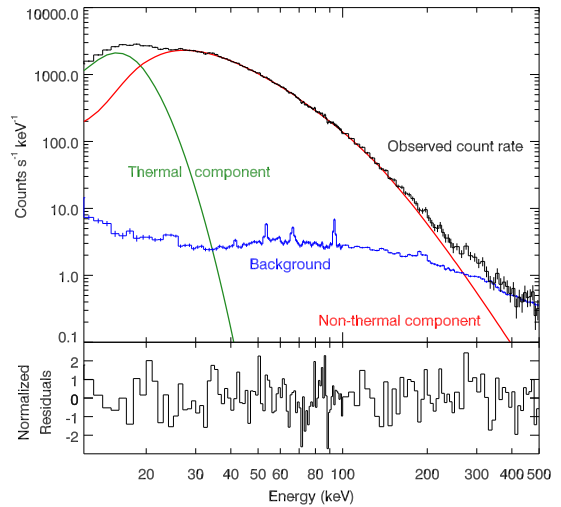}
\caption{
  Example of count spectrum ({\it thick error bars}) observed by RHESSI,
  integrated for 4 seconds around 17-JAN-2005 09:43:36 UT. The spectrum was
  fitted with a thermal component (Maxwellian), a non-thermal component
  (log-parabolic, see Appendix A), and the average spectrum before and after the
  flare, assumed to be the background. For clarity, the total model consisting
  of the sum of the three components is not plotted over the observed count
  spectra, but its normalized residuals are shown below instead. The residuals
  show that the fitted spectral model reproduces the observed counts. The reduced
  $\chi^2$ for this spectrum equals 0.94.   }
\label{fig:countspectrum}
\end{figure}

%
%

%
%
\begin{figure*}
\plotone{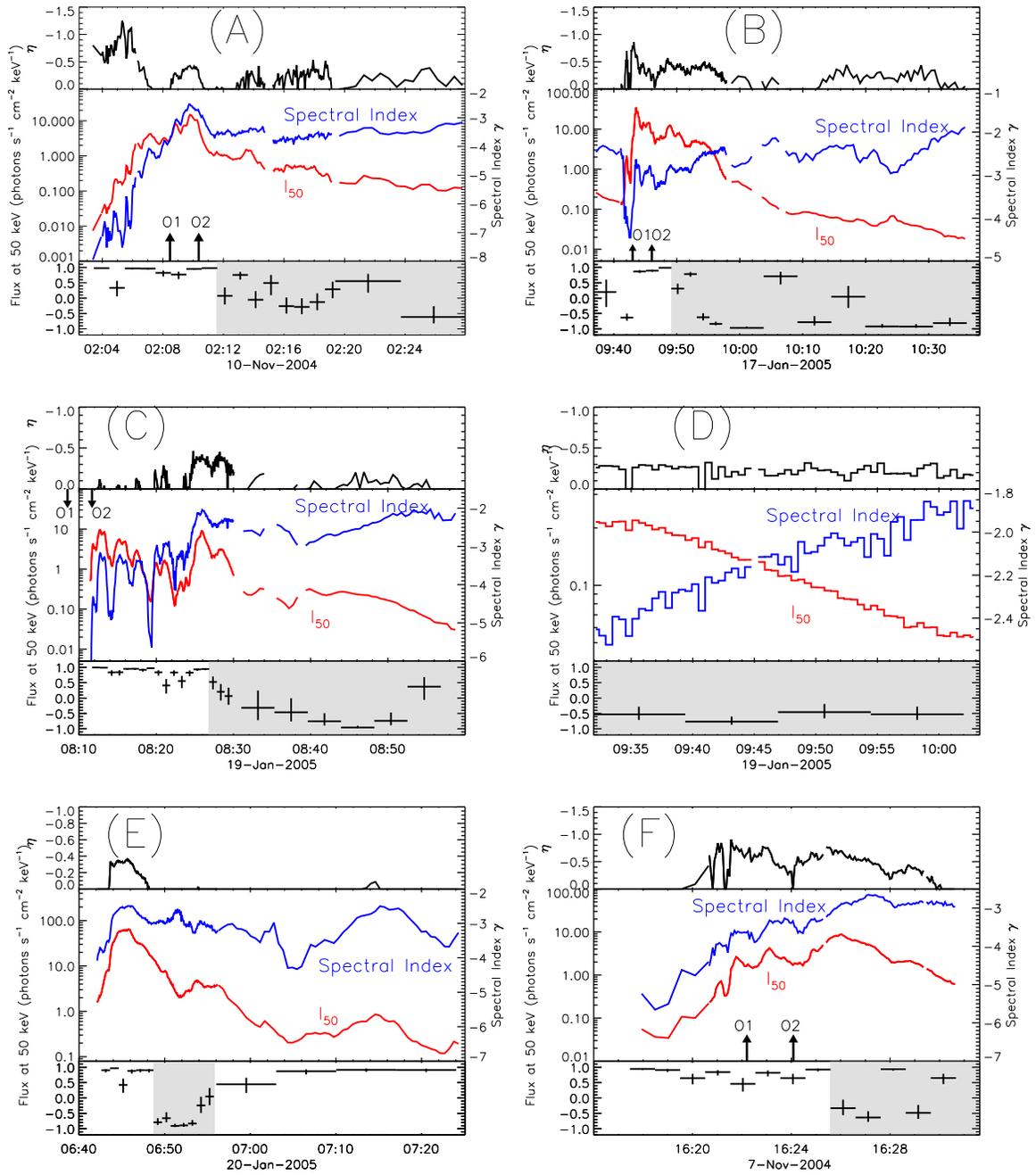}
\caption{
  Temporal evolutions of the spectral index $\gamma$, the flux at 50 keV,
  $I_{50}$, and the spectral curvature $\eta$ for the selected events. 
  Below, the correlation coefficient between
  the two curves is shown vs. time. It was determined during the time intervals
  indicated by the horizontal bars. The vertical bars represent the 68\%
  confidence range.  The onset times of the flare associated CMEs are marked by
  arrows.  O1 and O2 indicate linear and quadratic extrapolations, respectively,
  to the CME altitude vs. time evolution.}
\label{fig:lightcurves}
\end{figure*}
%
%

%
%
\begin{figure*}
\epsscale{0.6}
\plotone{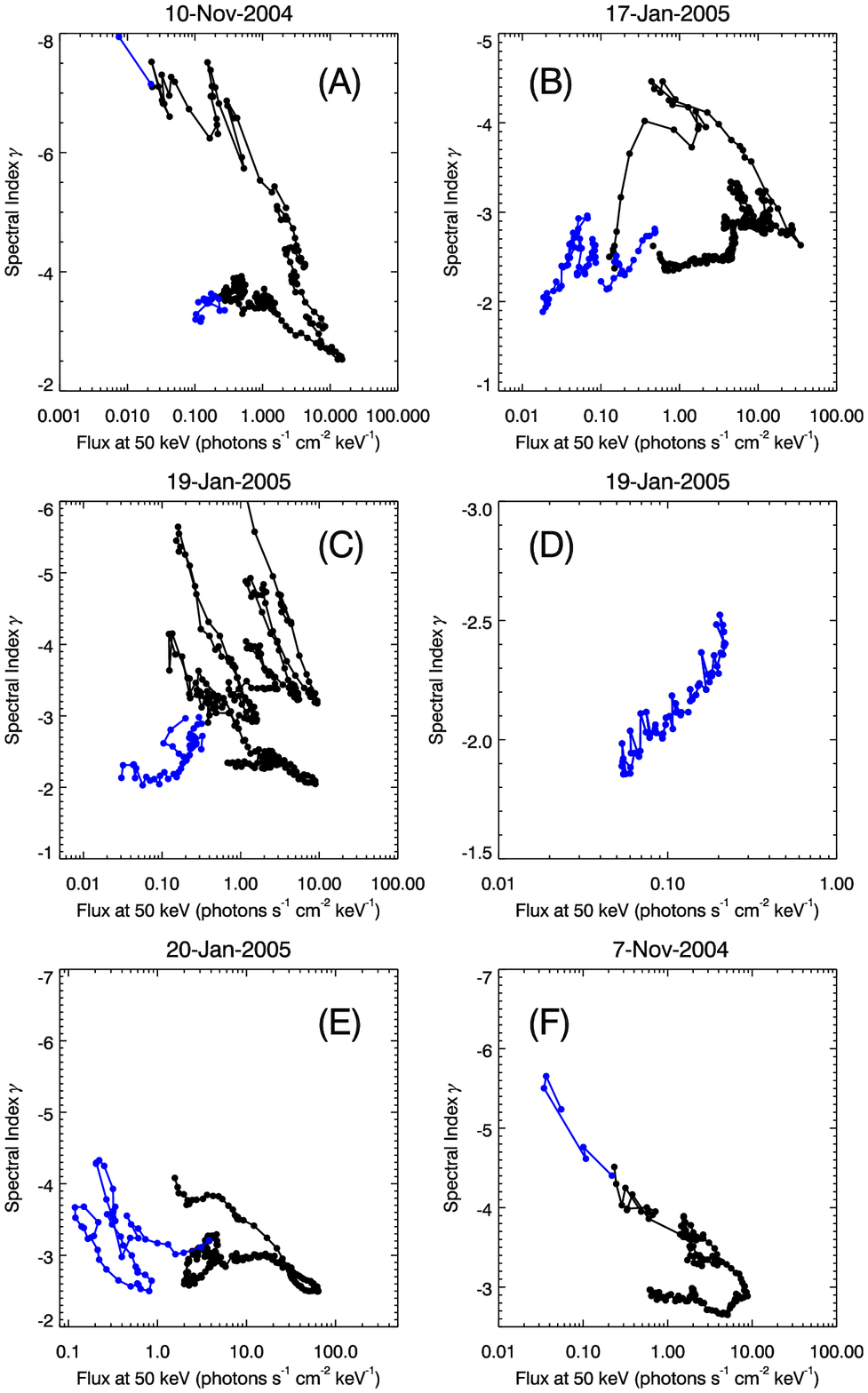}
\caption{Evolution of non-thermal X-ray spectral index $\gamma$ vs. photon flux
  $I_{50}$ in the course of flares.}
\label{fig:glogf}
\end{figure*}
 
%
%

The goal of this paper is a detailed quantitative study of the spectral
evolution of solar flares showing a hardening trend in RHESSI observations.
Rather than attempting a statistical study of a large number of flares, the
analysis is restricted to a few events studied exhaustively. Therefore, we do
not estimate the occurrence frequency of SHH flares or their rate of association
with solar energetic particle events. This has been done by \cite{kiplinger95}
using SMM/HXRBS data, who reports 24 occurrences of hardening out of 152 events
with peak flux count rate larger than 5000 counts s$^{-1}$. Most of the SHH
events reported by Kiplinger are in the upper M and X GOES class. The reality of
this trend needs however to be confirmed.

We selected flares with a GOES flux above X1 during RHESSI observation time
windows. 50 events satisfying this condition were found in the period from the
start of the mission (February 2002) to September 2006. We additionally required
that the rise, main and decay phases were well observed to study the spectral
evolution in time. This left us with 12 candidates.

As a first approximation, the presence of hardening behavior was tested by
studying the count rates in the energy range from 30 to 60 keV. We fitted the
spectral index separately in three energy bands (30--40 keV, 40--50 keV, 50--60
keV) and looked for either trends of progressive hardening or the presence of a
late hard phase. The lowest of these bands is sometimes contaminated by thermal
emission. This can be easily spotted by comparing the time profile with the
other bands. Using count rates is adequate to identify candidate events for
spectral hardening, but may have missed some events where hardening happens in a
phase of low flux close to the background. After discarding two further events
with high pileup, we found 5 well-observed events with a clear signature of
hardening. These are listed in Table \ref{tab:evlist2}.

For each of the selected events, the instrumental response matrices and
count-rate spectrograms covering the energy range from 3 to 500 keV were
generated for the front segments with a temporal resolution of one RHESSI spin
period (approximately 4 s). The spatially integrated photon spectra in the range
12 to 500 keV were fitted with two components: an isothermal component at low
energies (below about 20-40 keV) and a non-thermal component at higher
energies. The shape of the non-thermal component is assumed to be a
log-parabolic curve characterized by three parameters: its normalization
$I_{\eps_0}$ at energy $\eps_0$, its spectral index $\gamma <0$ and its spectral
curvature $\eta$.  The functional form of this model (Eq. \ref{eq:spmodel}), as
well as the rationale for choosing it are explained in detail in
App. \ref{app:a}. Figure \ref{fig:countspectrum} shows an example of an observed
count spectrum and the best-fit components. The normalized residuals
indicate an excellent fit.

During the fitting process, the model photon spectrum is folded with the
response matrix, yielding the expected count spectrum from the model. The
background spectrum is then multiplied with a normalization factor $\lambda$ and
added to the model counts, where $\lambda$ is an additional fit parameter for
the model fitting and is constrained between 0.5 and 2. This correspond
approximately to the maximum excursion in RHESSI's background in the front
segments during an orbit. The parameter $\eta$ is constrained to be zero or
positive (corresponding to a parabola in log-log bending down). This ensures
that the emission approaches 0 for infinitely high energies.

The large amount of data (more than 3 thousand spectra) required an automated
fitting routine. For every spectrum, 2 preliminary passes were done estimating
the parameters for the thermal and the non-thermal part which were then used as
starting parameters for the final fitting. This turned out to deliver good
fittings for most of the data. A check of the quality of the fits was performed
by looking at the time evolution of $\chi^2$ and of the fitting
parameters. Spectra with reduced $\chi^2$ worse than 2 were manually fitted
again, and in most cases it was possible to find another set of fit parameters
yielding reduced $\chi^2$ below 2, with the exception of the event of
20-JAN-2005.

This event (the largest flare, GOES class X7) is characterized by very strong
thermal emission. At times when the non-thermal emission is weak and/or soft,
pileup effects are especially large in the 20-50 keV band. Therefore, the
fittings, which are good above 50 keV, have large residuals below that
energy. This may be due to the fact that the photon spectrum model chosen is not
suited to describe the observed photon spectra, or that the pileup correction is
inaccurate. Because it is very hard to correctly take into account pileup
effects in such a regime, it is not clear whether the model failure is real or
instrumental. Therefore, we let the spectrum model stand as it is, but caution
that the parameter values fitted in the flare of 20-JAN-2005 may be
less accurate, due to the unknown systematic effects generated by imperfect
pileup correction.

For the other events, the fit parameters are of good quality and the
corresponding photon models are a high-fidelity representation of the incoming
photon flux. The final distribution of the reduced $\chi^2$ for all events
(except 20-JAN-2007) is such that 89\% of all spectra have $\chi^2$ less than
1.5 and 97\% of all spectra have $\chi^2$ less than 2. Therefore the unusual
choice of the logarithmic parabolic fit-model, explained in Sect. \ref{methods},
produces good fittings and is therefore justified \emph{a posteriori}.

Three of the events presented here have also been studied by
\cite{2008ApJ...673.1169S}, where the non-thermal component has been fitted by a
double power-law. The temporal evolution of our values for $\gamma$ (at 50 keV)
is very similar to theirs, although the actual numerical values are slightly
different due to the different fitting methods employed and their choice of
energy intervals. The differences are larger for the January 20 event, where
they used the 100-200 keV energy interval.

The thermal evolution of the events shows a rapid increase of the emission
measure at the beginning of the event, followed by a flat peak and a slow
decay. The temperature is in the range 20-40 MK, peaks before the emission
measure and decays faster.


\section{Spectroscopy Results}
\label{results}

Figure \ref{fig:lightcurves} shows time profiles of the photon spectral
index~$\gamma$ and the flux normalization at 50 keV,~$I_{50}$ (given in
Eq.~\ref{eq:spmodel}), for the events of Table~\ref{tab:evlist2}, as found by the
spectral fitting procedure explained in Sect.~\ref{methods} and Appendix A.

%
%
 
\begin{figure}
\plotone{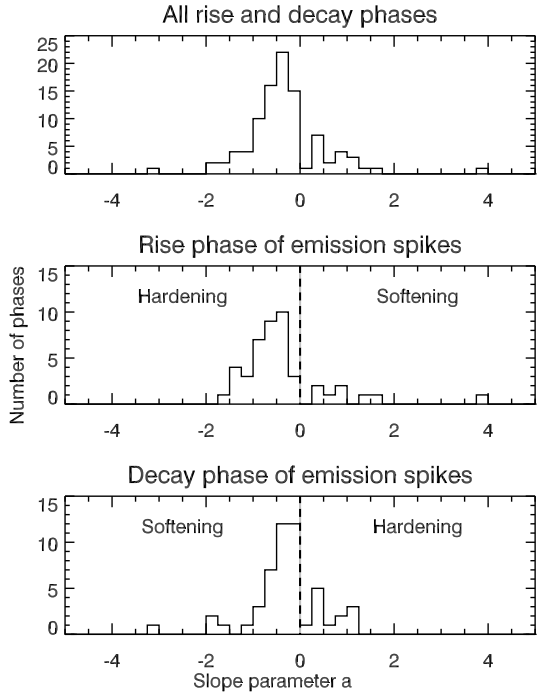}
\caption{Distribution of the slope in $\gamma$ vs. log($I_{50}$)
  (Fig. \ref{fig:glogf}, parameter $a$ in Eqs. \ref{eq:lintrendglogf} and
  \ref{eq:pivslope}) for the rise and decay phases in all peaks of the
  observed events (upper panel) and for rise and decay phases separately (middle
  and bottom panel, respectively).}
\label{fig:histoslope}
\end{figure}
 
%
%

The observed spectral variability of flares on time scales down to ten seconds
or less requires the highest possible temporal resolution for the spectral
fitting (about 4 seconds in our case). Longer integration times, while desirable
for better photon statistics, are not suitable because the averaging effect of
summing spectra with different hardness blurs the spectral evolution.
Nevertheless, during the decay phase the flux is so low that full resolution
spectra deliver noisy values for the fitting parameters. In this case, longer
integration times must be used. This is acceptable, since the variations of the
hard X-ray flux are slower during the decay phase, and short-lived spikes are
less frequent. Therefore, some light curves shown in Fig. \ref{fig:lightcurves}
and subsequent figures use a lower cadence of approximately 32 seconds (that is,
8 RHESSI rotations) in the decay phase.

Short gaps lasting about 1 minute can be seen in the light curves
(Fig. \ref{fig:lightcurves}). They correspond to periods where the thick
attenuator was removed from the field of view, but the X-ray flux was still so
large that the dead time in the detector prevents meaningful spectral analysis.

The selected flares show many distinct SHS peaks. They are characterized by a
temporal correlation of $\gamma$ and $\log I_{50}$, yielding roughly parallel
curves in Fig. \ref{fig:lightcurves}. On the other hand, in the presence of SHH
peaks or progressive hardenings, the two time profiles diverge. This may be
illustrated in the event of 19-JAN-2005 (Fig. \ref{fig:lightcurves}, panel C),
where the two profiles run roughly parallel until 08:29 and then start to
diverge.

To better distinguish between the SHS and the hardening trends,
Fig.~\ref{fig:lightcurves} also shows the correlation coefficients between the
spectral index $\gamma$ and the logarithm of the flux $\log I_{50}$ vs.
time. The vertical bars represent the 68\% confidence range (corresponding to
one standard deviation). SHS peaks are characterized by a correlation
coefficient close to +1. Times, on the other hand, when the spectrum hardens
while the flux becomes lower have a negative value of the correlation
coefficient. We note that during the SHS times correlation is rather constant
(near +1), whereas during the hardening phase the correlation coefficients are
more erratic and often not close to -1. This indicates that it may not be
possible to find a behavior similarly well-defined for the periods of hardening
as it is found for the SHS peaks. In some cases (19-JAN-2005, 08:28 to 08:30),
the spectral hardness stays nearly constant while the flux decays. This yields a
correlation coefficient near zero.

It should be noted here that most periods showing hardening have low
flux. Therefore the corresponding time profiles may be more noisy, weakening the
correlation. Some of this effect was compensated by increasing the time interval
for correlation. Another effect is loss of correlation during a broad
peak. Again this can be taken into account by increasing the correlation
interval.

During the late phase of the event of 19-JAN-2005, in the RHESSI orbit following
the one featuring the main peak, an uninterrupted phase of hardening is seen.
From 09:35 to 10:00 the flux decays exponentially (see
Fig. \ref{fig:lightcurves}, panel D). After that time, the emission reaches a
hardness comparable with the one of the background, and it becomes impossible to
disentangle the two components by purely spectral methods. This event will be
investigated in more detail in Section \ref{modeling}.

We also compared the start of the hardening with the onset time of
flare-associated CMEs, taken from the SOHO LASCO CME catalog
\citep{2004JGRA..10907105Y}. In three cases (panel A, B, and F in
Fig. \ref{fig:lightcurves}), the onset of the CME precedes the start of the
hardening by 3 to 5 minutes, in one case by 15--20 minutes (panel C) and in one
case by 50 minutes (panel E). In all five events an associated CME was
present, but the hardening phase never starts before the CME onset.

Figure \ref{fig:glogf} shows the relationship between $\gamma$ and $\log
I_{50}$. Since the flux increases from left to right and the hardness increases
from top to bottom, SHS peaks show as piecewise linear trends with a negative
slope \citep{grigis04,grigis05}. On the other hand, progressive hardening during
flux decay is visible as a trend with positive slope. Such relations can
be written as
\begin{equation}
\label{eq:lintrendglogf}
  \gamma=-a\log(I_{\eps_0})+b\,,
\end{equation}
where $a<0$ for the SHS peaks and $a>0$ during hardening phases. The minus
sign in front of $a$ takes care of the fact that in Fig. \ref{fig:glogf} the
vertical axis for $\gamma$ is reversed.

Careful examination of the plots in Fig. \ref{fig:lightcurves} and
\ref{fig:glogf} reveals that
\begin{itemize}
\item Most of the emission spikes are well represented by straight lines in
  $\gamma$-$\log I_{50}$ during the rise and decay phase. The decay phases are
  sometimes flatter in $\gamma$-$\log I_{50}$ than the corresponding rise phases
  (e.g. panel C), but the opposite is also observed (panel A). The event shown
  in panel E shows some significant deviations from the piecewise straight
  trend.
\item Spectral variability is stronger at the beginning of the event (panels A,
  B, C, F).
\item In the late phase of the events a slower varying component is seen,
  piecewise straight in $\gamma$-$\log I_{50}$, mostly nearly flat, slowly
  hardening (panels B, D), slowly softening (panels A, F), staying at an
  approximate constant hardness (panel C), or a mixture of the above (panel E).
\item During the rise phase up to the strongest peak, the hardness tends to
  increase from peak to peak (panels A, C). Events are softer in the beginning
  (all panels).
\end{itemize}

Figure \ref{fig:histoslope} shows the distribution of the $\gamma$-$\log I_{50}$ slopes
(parameter $a$ in Eqs. \ref{eq:lintrendglogf} and \ref{eq:pivslope}) in
the rise and decay phases of all emission spikes in the 5
events. If the flux increases, a negative slope represents spectral hardening and a
positive slope a softening. The opposite happens during a decay phase. SHS peaks
have $a<0$ both in the rise and decay phase.

The separate histograms for rise and decay phase differ marginally. Spectral
hardenings during rise have a slightly steeper $\gamma$-$\log I_{50}$ slope than
softenings during decays (SHS peaks). Quantitatively, the average value of $a$
when restricted to negative values is of $-0.70 \pm 0.06$ (the uncertainty is
the standard error of the mean) during the rise phases and $-0.51 \pm 0.08$
during the decay phases (after removing an outlier with slope -3.2). The average
of $a$ for the combined set of rise and decays is $-0.60 \pm 0.05$,
corresponding to an average pivot point energy (see App. B) of $\eps_*=9.4 \pm
1.3$ in agreement with \cite{grigis04}.

Examination of spectral hardenings during decay and softenings during rise
phases (trends opposite to SHS), restricted to the range 0--3, yield an average
$a$ value of 0.66$\pm$0.11 and 0.88$\pm$0.18, respectively. These values are not
significantly different from each other or from the corresponding absolute value
of the $a<0$ averages.

\section{Imaging results}

%
%

\begin{table*}
{\scriptsize 
\caption{Source motion near the onset of hardening.}

\label{tab:hardeffect}
\centering
\begin{tabular}{l l l l}
\hline\hline
Event & Onset of  & Footpoints  & Coronal source \\
      & hardening & motion      & motion         \\
\hline 
07-NOV-2004 & 16:26 & Jump in the position of the eastern FP 
                    & uncertain (bad images at\\
            &       & \ \ Change in the direction of motion of the western FP 
                    & \ \ low energies after onset) \\ 
10-NOV-2004 & 02:12 & Jump in the position of the two brighter FPs & stationary   \\
17-JAN-2005 & 09:49 & Nearly continuous motion of both FPs & continuous motion \\
19-JAN-2005 & 08:27 & Continuous motion of the northern FP, slowing after the onset &
continuous motion upward  \\
            &       & Nearly stationary position of the southern FP & \\
20-JAN-2005 & 06:49 & Reversal in the direction of motion of both FPs, slowing down
afterwards & continuous motion upward\\
\end{tabular}
}
\end{table*}
 
%

Is there a connection between spectral hardening and X-ray source geometry
indicating a different acceleration site? The position of the hard X-ray
footpoint sources were investigated in CLEAN images, using detectors 3 to 8 with
a cadence of 60 seconds. In particular, we looked for differences in source
positions and velocities at the onset and during the period of general
hardening.

\begin{figure*}
\epsscale{0.55}
\begin{tabular}{ccc}
\plotone{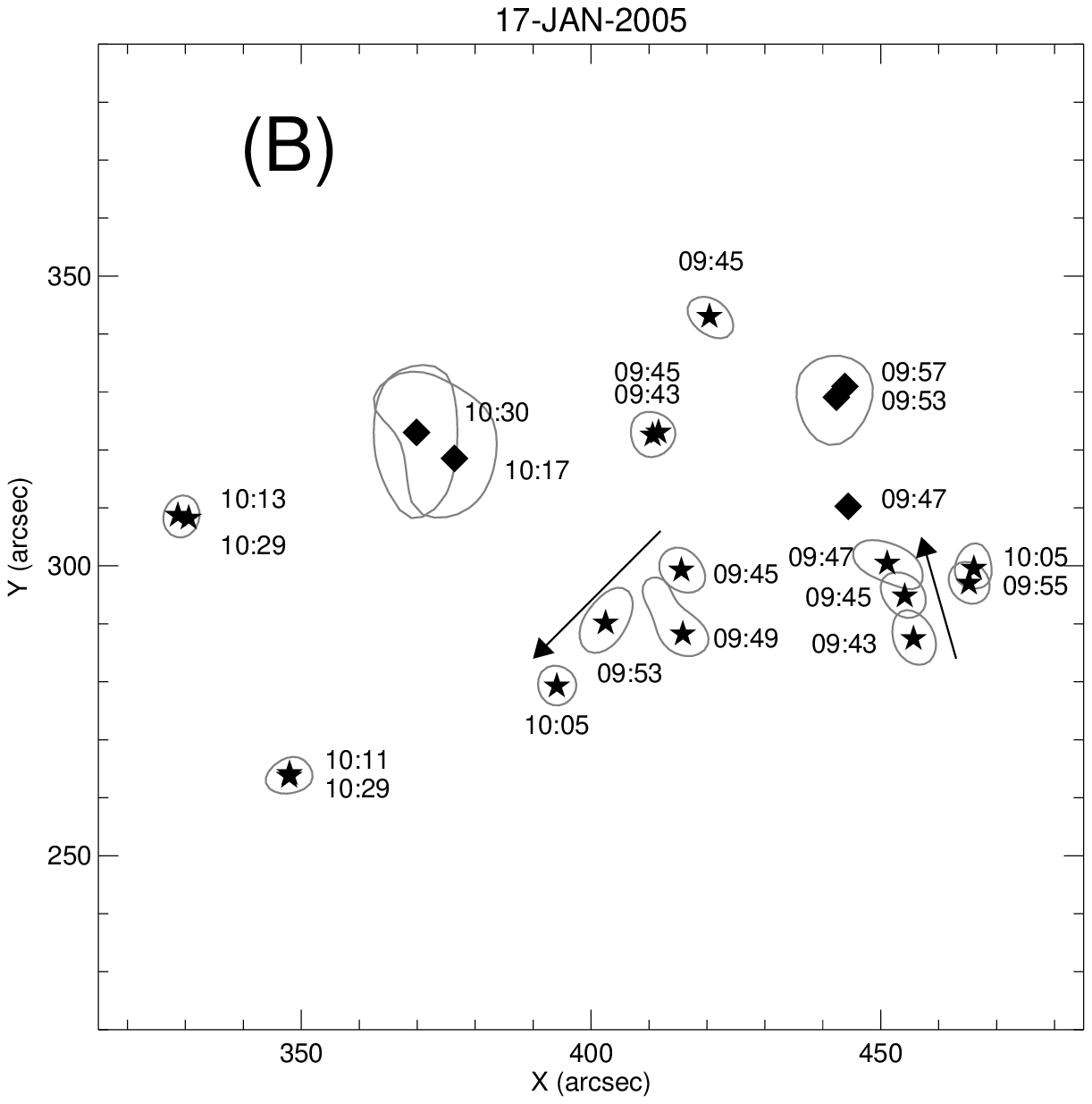} & \hspace{-2cm}\plotone{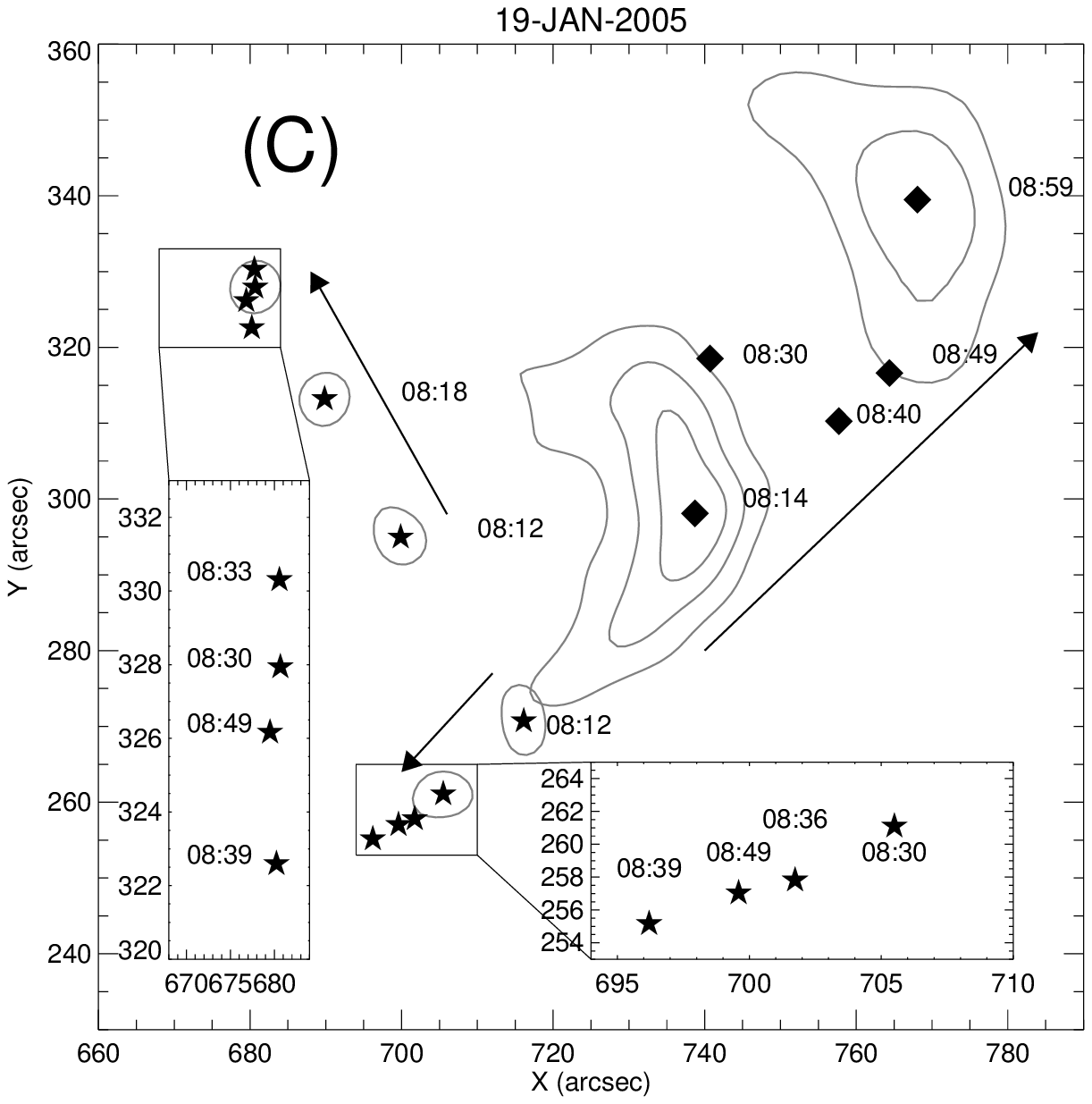} & \\
\plotone{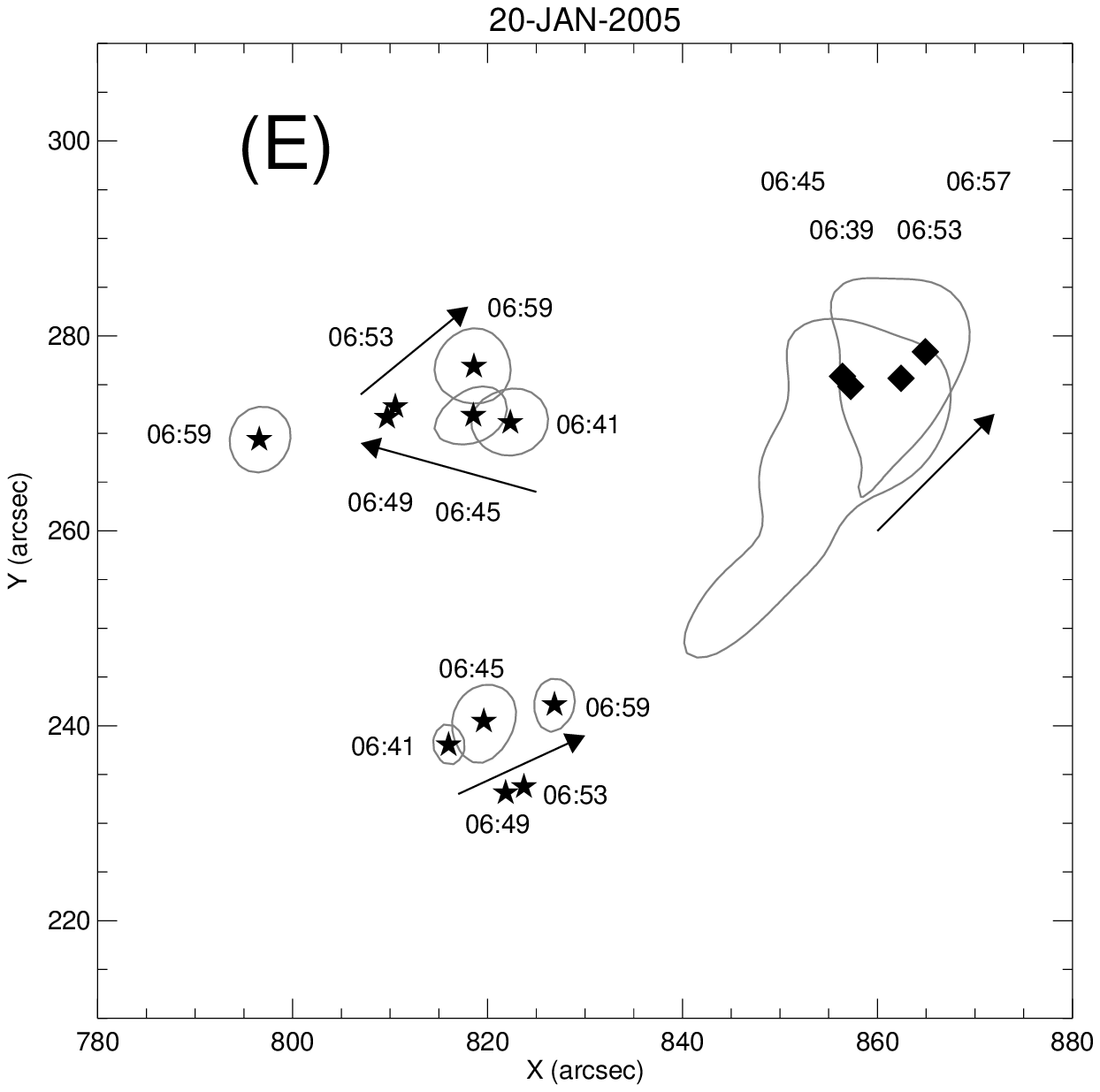} & \hspace{-2cm}\plotone{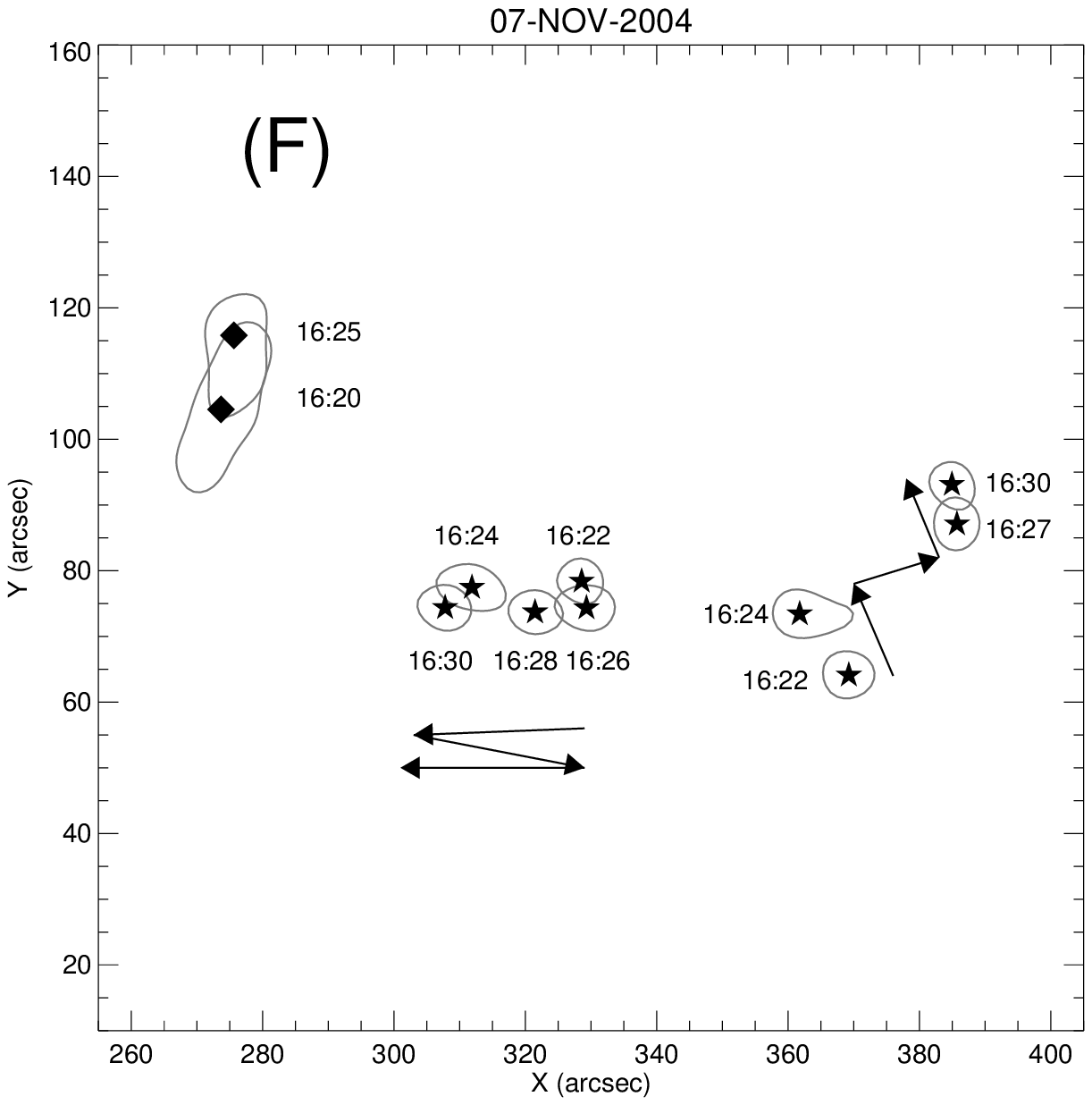} & \\
\end{tabular}
\caption{ RHESSI hard X-ray positions of the non-thermal footpoint sources
  (stars) and the thermal coronal source (diamond) for 4 events. The energy
  range for the footpoint position is 50-100 keV band (or 25-50 keV at times
  where the 50-100 keV emission is too faint for imaging). The energy range for
  the coronal source position is 12-25 keV. Systematic motions are indicated by arrows.}
\label{fig:evc}
\end{figure*}

Figure \ref{fig:evc} shows an overview of the position of the thermal
and non-thermal sources during the events. The event of 10-NOV-2004 is not shown
because it lacks a coordinated evolution of source positions.

The event of 19-JAN-2005 is particularly interesting, as it was well observed
and the footpoints (FP) clearly move along the ribbons noticeable in a TRACE
image at 1600 \r{A} \citep{2008ApJ...673.1169S}. The motion is fast at the
beginning and later slows down. The thermal source has the form of a loop,
rising throughout the event.

\begin{figure}
\plotone{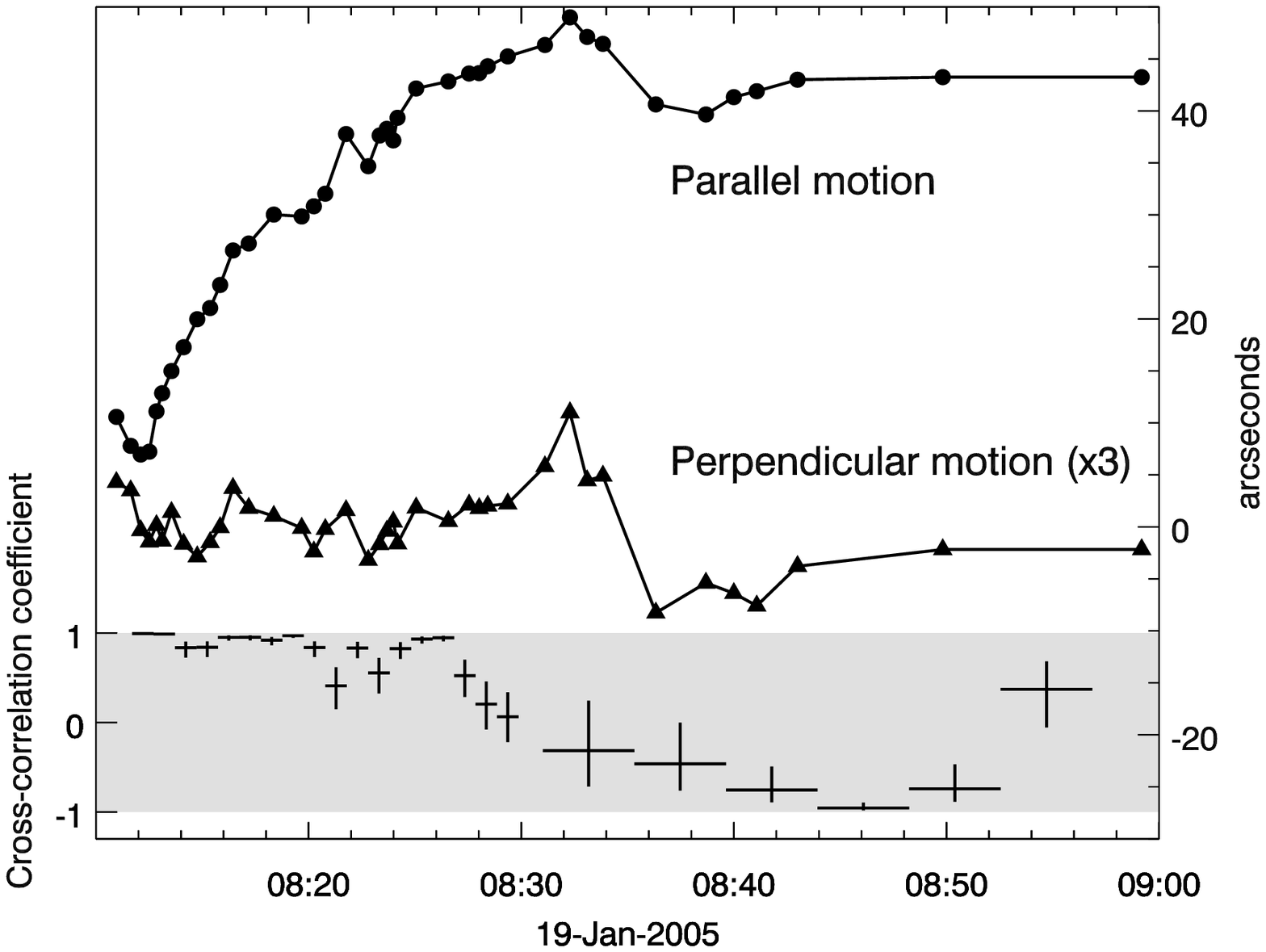}
\caption{Movement of the northern X-ray footpoint source for the event of
  19-JAN-2006. The upper curves show the motion in the two components parallel
  (top curve) and perpendicular (middle curve, multiplied by a factor of 3) to
  the northern TRACE 1600 \r{A} ribbon. The bottom curve displays the cross
  correlation coefficient between hard X-ray spectral index and flux, showing
  the onset of spectral hardening around 08:27.}
\label{fig:fpmov_ccross}
\end{figure}

Figure \ref{fig:fpmov_ccross} shows the displacement of the northern FP source
in the directions parallel and perpendicular to the northern ribbon\footnote{The
  parallel and perpendicular directions are defined relative to the direction of
  the regression line obtained by least-square fitting the positions of the
  hard X-ray footpoints independently in each ribbon.}. The FP source starts with
a velocity of 50 km s$^{-1}$ along the ribbon and slows down continuously during
the SHS phase until the onset of the hardening phase, when it becomes nearly
stationary. There is no evidence of an abrupt transition between the two
regimes. Furthermore, 19-JAN-2005 is the only event clearly showing footpoints
drifting apart and coming to a stop at the onset of hardening.

The other events show other positional changes. In the following the geometrical
behavior of the sources and their evolution are described shortly for all the
events. We distinguish between footpoint sources and coronal sources. Because of
projection effects, we cannot reliably reconstruct the three dimensional
structure of the sources. However we know from limb event observations
\citep{battaglia06} that FP sources are mainly non-thermal and well-observed
above 20-30 keV, whereas coronal sources are mainly thermal and well-observed
below 20-30 keV. Therefore, we assume in the following that the thermal source
is coronal and that non-thermal sources are footpoints.

{\bf 07-NOV-2004:} This events features two footpoints and a coronal source.
The eastern FP moves from W to E from 16:20 to 16:24 UT, jumps back to near the starting
position and moves again from W to E from 16:25 to 16:30. The western FP moves
slightly from SW to NE from 16:21 to 16:24, then changes direction with a slight 
jump to W, and slowly moves to NW from 16:25 to 16:30. The western FP is brighter
than the eastern FP before 16:24 and dimmer after 16:25.  The thermal source is
located farther E than the FPs and moves slightly to N from 16:19 to 16:25, and is
not clearly seen in the images afterwards, due to the insertion of RHESSI's
thick attenuators.

The jump in position around 16:25 roughly coincides with the time at which the
hardening starts, possibly indicating that another loop is actively accelerating
electrons, but the expansion of the loop, as suggested by the FP motions,
continues during the hardening phase, contrarily to what has been observed in
the event of 19-JAN-2005.\\

{\bf 10-NOV-2004:} This event has a very complicated FP morphology, with
sources and source-pairs appearing in many different places. It is not possible
to find a well-defined source motion like in the simpler cases with only two
footpoints. Here the sources seem to jump around as new footpoints in a
different position become brighter and outshine the old ones. Important shifts
in position occur at 02:08 and 02:12. The latter shift happens at the same time
as the onset of hardening.\\

{\bf 17-JAN-2005:} Three pairs of FPs are seen. The northern pair is stationary
and seen from 09:43 to 09:45. The southern pair consists of an eastern FP moving
to SE from 09:45 to 10:05, and a western FP moving to N from 09:43 to 09:47, then
shifting to W (09:55) and moving very slowly to N until 10:05. The last pair of FP
is to the east and stationary from 10:11 to 10:29.  Coronal sources are seen in
two locations: one to the N of the southern FP pair, moving to N from 09:46 to
09:57, and the second to the NW of the easternmost FP pair, nearly stationary
from 10:16 to 10:30.

There is no clear signature of a discontinuity or a change of behavior
happening around 09:50, when the hardening starts.\\

{\bf 19-JAN-2005:} Two footpoints are seen with a loop-shaped coronal source
between them. The northern FP moves to NE from 08:12 to 08:30 (covering nearly 60
arcseconds) while the southern FP moves, slower, to SE. In the meantime, the
loop-shaped coronal source moves to NW (indicating that it is 
rising). After 08:30 (only 3 minutes after the onset of hardening) the northern
FP is much slower. The coronal source keeps moving to NNW. In the next RHESSI
orbit (after 09:30), the FP sources can still be seen near the old positions at
08:30. The northern FP is nearly stationary from 09:33 to 09:59, while the southern
FP very slightly moves to W, and the coronal source slightly moves to N.\\

%
%
 
{\bf 20-JAN-2005:} This near-limb event features two FPs and a loop-like coronal
source. The southern FP moves to W while the northern FP moves to E. The
eastward motion of the northern FP is not continuous: at 06:49 it reverses and
recedes until 06:55, when another sources appears 20$^{\prime\prime}$ to E. The
double structure lasts until 07:01, when the easternmost source fades away. The
coronal source moves to NW and rises throughout the event, slowing down towards
the end. The reversal coincides with the start of the hardening phase.\\

The observed behavior at the onset of hardening for all events is reported in
Table \ref{tab:hardeffect}. It summarizes the analysis of the source motions in
an interval of time spanning 4 minutes, centered on the onset of hardening. We
conclude from the imaging observations that there is no universal trend
holding for all events. Sometimes, there seems to be a switch to a different
loop system near the beginning of the hardening phase.  On the other hand, such
jumps can also be seen during the SHS phase of the events, so they need not be
significant. There is also some indication that the FP motion is slower during
the hardening phase, but again this does not hold for all events. In the event
of 19-JAN-2005, with a simple geometry and well observed, the change in spectral
behavior leading to the hardening phase does not have an impact on the
morphology of the hard X-ray sources seen by RHESSI.

%
%
 
%
%

\section{Modeling the SHH phase}
\label{modeling}

\cite{grigis06} showed that the soft-hard-soft trend is expected from a
transit-time damping stochastic acceleration model that includes escape of
particles from the accelerator. The hardness is controlled by how fast the
particle gain energy and how long they are trapped in the accelerator. Harder
spectra result from longer dwelling times of the electrons in the accelerator
and higher acceleration efficiency. These conditions also allow a larger
population of high-energy electrons to build up, leading to increased hard X-ray
emission from the accelerator, identified as a part of the looptop source.  This
basic model predicts that harder spectra also have larger hard X-ray flux, but
cannot explain the soft-hard-harder trend seen as the flux decays, because 
  these observations associate harder spectra with smaller flux.

To fit the observed SHS behavior, \cite{grigis06} had to assume that electrons
are trapped below a certain threshold energy $E_\mathrm{T}$ and cannot leave the
accelerator. The escaping electron population has a low-energy cutoff at
$E_\mathrm{T}$. Then the photon spectra \emph{of the footpoints}, which dominate
the non-thermal emission, harden below $E_\mathrm{T}$.

In the following, a simple extension of the basic stochastic acceleration model
is presented which could lead to the observed spectral hardening. We introduce
the additional assumption that $E_\mathrm{T}$ increases with time in the SHH
phase. Therefore, the photon spectrum below $E_\mathrm{T}$ hardens while at the
same time the flux arriving at the footpoints decreases. The important point
here is that this also happens if the electron spectral index $\delta$ and the
electron flux normalization in the acceleration region are constant. Thus the
new variable does not contradict the basic properties of the stochastic
acceleration model. This is shown in Fig. \ref{fig:spmod}, where footpoint
photon spectra have been computed from a given electron distribution with
various $E_\mathrm{T}$. As expected, the photon flux decreases with increasing
$E_\mathrm{T}$. Note that the photon spectra have a nearly constant but slightly
increasing hardness while the flux decreases.

This extension is compared numerically to the late phase of the 19-JAN-2005
event. For simplicity, an electron distribution with constant spectral index and
flux normalization is assumed to escape from the acceleration region. It has a
low-energy cutoff $E_\mathrm{T}$ due to trapping in the
accelerator. $E_\mathrm{T}$ increases with time. Obviously, we do not expect
such a simple scenario to reproduce all the details of the spectral
evolution. The question is, how much of the observed features can be explained
with this simplest extension of the existing acceleration model.

A non-isotropic energy distribution of fast electrons with a positive slope in
energy is known to be unstable towards growing plasma waves. Therefore our
scenario also includes an alternative electron distribution featuring a turnover
at $E_{T}$, that is, a flat distribution below $E_{T}$, instead of a
cutoff. The electron distributions with cutoff and turnover are assumed as,

\begin{eqnarray}
\label{eqn:modelsp}
  F_\mathrm{CUTOFF}(E) & = & \left\{
       \begin{array}{lll}
         F_{E_0}\left(\displaystyle\frac{E}{E_0}\right)^{\delta} 
                              & \mathrm{if} & E\ge E_\mathrm{T}\\
         0 &\mathrm{if} & E< E_\mathrm{T} \\
         \end{array}\right.\\
  F_\mathrm{TURNOVER}(E) & = &  \left\{ \begin{array}{lll}        
    F_{E_0}\left(\displaystyle\frac{E}{E_0}\right)^{\delta} 
    & \mathrm{if} & E\ge E_\mathrm{T}\\
    F_{E_0}\left(\displaystyle\frac{E_{T}}{E_0}\right)^{\delta} &\mathrm{if} & E< E_\mathrm{T}
         \end{array}\right.
\end{eqnarray}

The free parameters are the electron spectral index $\delta$, the electron flux
normalization $F_{E_0}$ (electrons s$^{-1}$ keV$^{-1}$), the cutoff or turnover
energy $E_\mathrm{T}$. The reference energy $E_0$ is fixed at 50~keV.

An exhaustive description of the method used for the comparison of this
simple model with the observation and finding the best fit is given in
App. \ref{app:b}.

\begin{figure}
\plotone{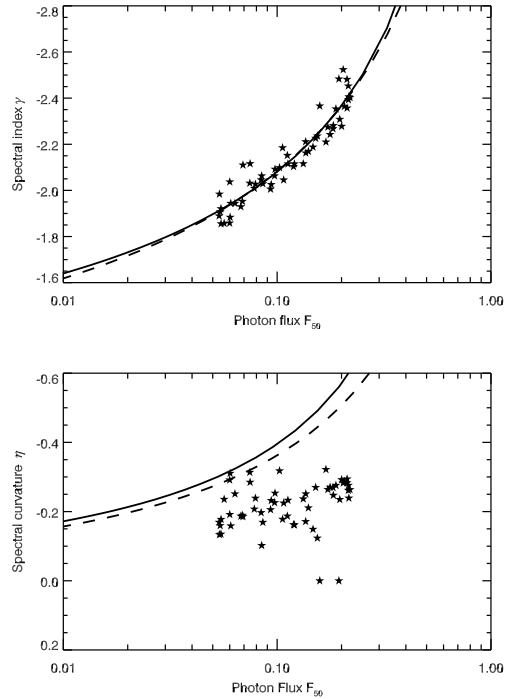}
\caption{ 
  Observed values of $\gamma$ vs. $I_{50}$ (\emph{top panel}) and $\eta$
  vs. $I_{50}$ (\emph{bottom panel}), represented by asterisks. The 
    best-fit model curves (continuous for the cutoff model, dashed for the
  turnover model) are expected from model with constant electron hardness and
  flux, but rising low energy cutoff or turnover energy $E_\mathrm{T}$.}
\label{fig:vdc}
\end{figure}

The late phase of the event of 19-JAN-2005, from 09:32 to 10:02, comprises 30
minutes of continuous hardening, and therefore is well suited for the comparison
with the variable cutoff model. Figure \ref{fig:vdc} shows the comparison. The
observed values of $\gamma$ and $\eta$ are plotted as a function of $I_{50}$
together with the best-fit model curve for both the cutoff and turnover electron
spectra. For the cutoff model, the best fit values of the parameters are:
spectral index $\delta= 6.52$, flux normalization at 50 keV $F_{50}=4.50\cdot
10^{33}$ electrons s$^{-1}$ keV$^{-1}$. In the course of the decay,
$E_\mathrm{T}$ increases from 98 keV to 159 keV (thus yielding a photon spectral
index $\gamma$ in the range between -2.5 and -1.9). The corresponding values for
the turnover model are: $\delta=6.11$, $F_{50}=3.67\cdot 10^{33}$ electrons
s$^{-1}$ keV$^{-1}$, while $E_\mathrm{T}$ increases from 127 to 224 keV.

The photon spectra from both the cutoff and turnover model are curved downward
in the fitted energy range. Thus the spectral curvature, $\eta$, is
negative. It is observed to be between 0 and -0.3, whereas the model spectra
have values in the range from -0.55 to -0.25 (Fig. \ref{fig:vdc}, bottom)
and thus are significantly more curved.

In the cutoff distribution, the total fluxes of electrons escaping the
acceleration site are initially $9.9\cdot10^{32}$ electrons s$^{-1}$ above
$E_\mathrm{T}=98$ keV. This reduces to $6.6\cdot10^{31}$ electrons s$^{-1}$ for
$E_\mathrm{T}=159$ keV in the course of the decay. The total injected power
reduces from $1.9\cdot 10^{26}$ erg s$^{-1}$ to $2.1\cdot 10^{25}$ erg
s$^{-1}$. In the turnover model, the total number of particles diminishes from
$2.3\cdot 10^{33}$ to $1.3\cdot 10^{32}$, and the injected power from $1.9\cdot
10^{26}$ erg to $2.8\cdot 10^{25}$ erg.

\section{Discussion}
\label{discussion}

Spectroscopic RHESSI observations are well suited to study the spectral
evolution during the main phase of large flares.  The path observed in the
$\gamma$ vs. $\log I_{50}$ plots for the events is not simple. However, it can
be broken down reasonably into a superposition of linear sections during flux
rise and decay phases. While not all rise or decay phases can be so decomposed,
this simple description is adequate for most of them, and permits comparison of
observations and theory.

There is a difference between the results reported here and the results from
\cite{grigis04} in the asymmetry between rise and decay phases in SHS peaks. The
previous results indicated that decay phases are steeper in the $\gamma$
vs. $\log I_{50}$ plot than rise phases. We find the
opposite. The reason is probably the selection bias: here we selected 
specifically events showing hardening. This hardening trend sometimes
overlays SHS peaks, giving rise to a soft-hard-less-soft pattern.

The hard X-ray images during the events show the usual morphology of hard X-ray
solar flares: a low-energy coronal source and two or more high-energy
footpoint sources. The position of the footpoint sources is strongly variable:
 it either moves smoothly or jumps from location to location. This reflects
changes in the connection between the accelerator and the chromosphere, as well
as in the location of the accelerator itself.

The behavior observed in the images cannot be reduced to one simple scenario
valid for all events. However, the observations suggest that there is
no clear separation between the SHS and the hardening phases: the former seems
to smoothly merge into the latter. Even in the cases where the emission jumps at
the onset of hardening (Table \ref{tab:hardeffect}), the footpoint behavior
seems not to change radically.

%
%
 
\begin{figure}
\plotone{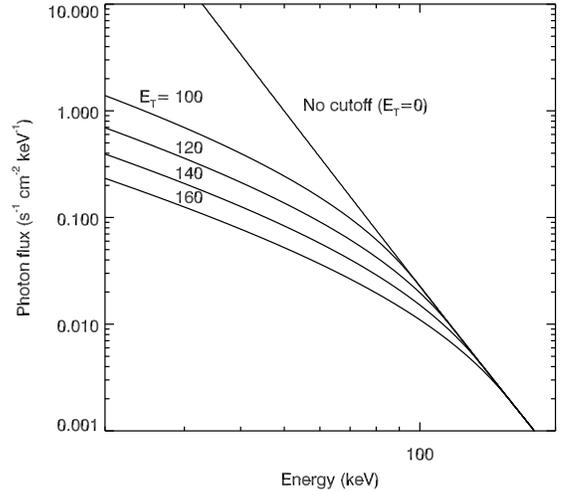}
\caption{ 
  Photon spectra by thick target emission (at footpoints) from a
  constant power-law electron distribution having spectral index of
  $\delta=-6.5$ in the acceleration region. The low-energy cutoff,
  $E_\mathrm{T}$, increases from the top to the bottom curve from 0, 100,
  120, 140, to 160 keV. 
}
\label{fig:spmod}
\end{figure}
 
%

Alternatives to the scenario presented in Sect. \ref{modeling} are
conceivable. In particular, electron storage in the corona and slow release
during the decay could be a possibility. As Coulomb interactions are faster at
low particle energies, the spectrum would harden with time while the released
flux decreases.  Noting that the hardening phase in the 19-JAN-2005 event lasts
more than 30 minutes and that the decay of the flux in time is nearly exponential
(as seen by the fact that the $I_{50}$ line in Fig. \ref{fig:lightcurves}, panel
D, is nearly straight), the total number of injected electrons can be computed
from the total electron fluxes at the start $F_\mathrm{BEG}$ and at the end
$F_\mathrm{END}$.
\begin{equation}
  F_\mathrm{TOT}=\Delta t
      \frac{F_\mathrm{END}-F_\mathrm{BEG}}{\log\displaystyle\frac{F_\mathrm{END}}{F_\mathrm{BEG}}}\,,
\end{equation}
where $\Delta t$ is the observed duration (here 30 minutes) and $F_\mathrm{TOT}$
the total injected flux. From the observed values, we get $F_\mathrm{TOT}\simeq
1\cdot 10^{36}$ electrons for the turnover model and $F_\mathrm{TOT}\simeq
6\cdot 10^{35}$ electrons for the cutoff model. These numbers do not seem
extraordinarily high, but it should be noted that all these electron have
energies above 100 keV.

Would such a large population of electrons be seen as a coronal hard X-ray
source in the 50 - 100 keV band? The luminosity depends on the volume and
density in the storage region. The observed footpoints in the hardening phase
are separated from each other by approximately 60$^{\prime\prime}$, indicating a
medium sized loop. Therefore, it would be visible on RHESSI images unless it
were excessively under-dense. Thus we reject the storage model.

Both the cutoff and turnover model are able to reproduce the observed $\gamma$
vs. $\log I_{50}$ trend, but fail to reproduce the correct spectral curvature
$\eta$ (Fig. \ref{fig:vdc}). Although the observation of $\eta$ is more difficult in the decay phase
due to the lower signal-to-background ratio, the difference between the cutoff
and turnover models and the observed points is significant. The value of the
parameter $\eta$ depends on the energy interval chosen for the fitting of the
model photon spectra (20-80 keV in our case). A lower maximum energy of this
interval produces lower model values for $\eta$.

We note furthermore that if the accelerator is inhomogeneous, the electron
spectrum at the footpoints is the superposition of different components with
different values of the low-energy cutoff or turnover $E_\mathrm{T}$. The
superposition of spectra that curve at different energies is less curved than
individual spectra. Figure \ref{fig:spmod} suggests that the superposition of
components with different spectral shape may in fact reduce of the total
spectral curvature. Thus we consider the disagreement in curvature not 
crucial to reject the model.

%
%

\section{Conclusions}
\label{conclusion}

We have presented results of spectroscopy, imaging and simple modeling of
spectral hardening observed occasionally in the hard X-ray emission of large
flares. The main conclusions are:

\begin{itemize}
\item The flares selected for the presence of a hardening phase also show
  soft-hard-soft behavior, at least initially. The hardening starts at or
  after the largest peak of the flares. In 3 out of 5 events it starts 2 to 6
  minutes after the onset of a CME.
\item Similar to SHS peaks, hardening phases can usually be described by
  piecewise linear sections in a plot of spectral index vs. logarithmic flux.
\item There is no clear trend relating the behavior of hard X-ray footpoint sources with
  the spectral evolution that would be valid for all events. Sometimes the location of the
  emission shifts when the hardening starts, in other events it does not.
\item In the event of 19-JAN-2005, there are only two well-defined
  footpoint sources throughout the whole event. No
  discontinuity is observed in the motion at the onset of hardening, but a general trend
  of slowing down, such that the FPs become nearly stationary during the decay phase.
\item In 3 out of 5 flares, the coronal source moved continuously during
  the onset of the hardening. This motion was directed upwards in
  two near-limb events.
\end{itemize}

%
%
 
%

In the sample studied, we find a surprising lack of detailed correlation between
spectral and spatial behavior. It is similar to what has been observed by
\cite{2005ApJ...625L.143G} in a smaller flare (M6) featuring strong
footpoint motions and hardening at the end.

The main question addressed in this paper is whether the SHS peaks and the
hardening phases are the results of two different acceleration mechanisms. The
results support the view that the same acceleration mechanism changes gradually
in the later phase of the flare. This change has clear effects on the spectrum,
but a more indirect and subtle influence on the source position. The operation
of a second acceleration process later in the flare cannot be ruled out,
however. Nevertheless, we have found strong evidence that there is a gradual
change in the accelerator, transforming its behavior from impulsive (showing up
as SHS peaks) to gradual (hardening phases). This is substantiated by the
observations of the superimposition of SHS peaks with a continuous hardening
trend and of the smooth footpoint motions during the onset of hardening.

The reason for the association with interplanetary proton events
\citep{kiplinger95} remains to be explored. As an aside we may note that when
the footpoints drift apart, acceleration takes place in larger and larger
loops. In a stochastic acceleration framework, the acceleration efficiency of
electrons in larger loops is reduced, while ions can be more efficiently
accelerated \citep{emslie04}. Since hardening trends are well correlated with
the occurrence of interplanetary energetic protons events, it is possible that
the very conditions that are responsible for the hardening trends favor
acceleration of protons, which may then escape into interplanetary space, with
the CME controlling their release rather than acceleration
\citep{2006A&A...445..715S}.

The observed motion of footpoints suggests that different coronal loops may be
involved in particle acceleration during a flare. They will have different
physical properties such as size, density, and magnetic field. The overall
magnetic geometry of the active region will determine which loops reconnect at
which time, sometimes giving rise to an orderly motion of footpoints, sometimes
generating a more chaotic situation. The data suggest that as the reconnection
process proceeds, some physical parameters of the acceleration site changes in
such a way as to favor the production of harder spectra, rather than having a
totally new process (say, shock acceleration) taking over in the decay phase.


\acknowledgements
  We thank S. Krucker, Y. Su and the participants of the 7th general RHESSI
  Workshop for useful discussions.  The analysis of RHESSI data at ETH Zurich is
  partially supported by the Swiss National Science Foundation (grant
  nr. 20-67995.02). This research has made use of NASA's Astrophysics Data
  System Bibliographic Services.  The SOHO LASCO CME catalog is generated and
  maintained at the CDAW Data Center by NASA and The Catholic University of
  America in cooperation with the Naval Research Laboratory.
  {\it Facilities:} \facility{RHESSI}

\appendix
\section{Hard X-ray model fitting}
\label{app:a}

The hard X-ray spectrum observed in solar flares consists of two distinct
components at low energies (that is, below 10-40 keV) and high energies. The
properties of the low-energy component are typical of thermal emission of a hot
plasma with temperature of 10 to 40 MK. In particular, the spectrum steepens
with energy, and the temporal evolution follows the (cooler) thermal plasma
observed in soft X-rays and EUV. 

The high-energy component behaves differently and is hence dubbed
non-thermal. It is harder then the thermal component and is usually fitted with
a power-law function of the energy with 2 free parameters. However, sometimes
the observed spectrum steepens at higher energies. In the literature, this is
usually accounted for by using a broken power-law model
\citep[e.g.][]{dulk92,battaglia05}. There are some disadvantages in the broken
power-law model: a) it is not physical, in the sense that any continuous
electron distribution emitting X-rays should generate a differentiable photon
spectrum, and b) the location of the break-point is poorly determined by the
observations.

We argue that there is a simpler extension to the power-law model which both
turns down at higher energies and is smooth. Recalling that a power-law function
plotted in log-log space is a straight line, we choose as a ``natural''
extension to the next order a parabolic model in log-log space, described by the
function
\begin{equation}
  \label{eq:spmodel}
  I(\eps)=I_{\eps_0}\cdot\left(\frac{\eps}{\eps_0}\right)^{\displaystyle \gamma -\eta\log\left(\eps/\eps_0\right)}\,.
\end{equation}
The 3 model parameters are the normalization $I_{\eps_0}$, amounting to the flux
at the (fixed) normalization energy $\eps_0$, the spectral index $\gamma<0$ and
the parabolic coefficient $\eta$, which we will refer to as the \emph{spectral
  curvature} although, strictly speaking, the geometric curvature of the
parabola is not constant, but equals $-2\eta$ in the vertex and vanishes at
infinity.

In the special case $\eta=0$, the (unbroken) power-law model is recovered. We
note here that a log-parabolic model has been used previously to describe
observed X-ray spectra of pulsars \citep{massaro00}.

In summary, the reasons for preferring the log-parabolic model over the more
usual broken power-law are:
\begin{enumerate}
\item It is simpler than the broken power-law, as it allows only 3 instead of 4
  free parameters.
\item For the vast majority of the time intervals, it produces similar values of
  $\chi^2$ as the broken-power law.
\item It is differentiable, therefore there exists a continuous electron
  spectrum producing the photon spectrum. This is not the case for the broken
  power-law, where a discontinuity is needed in the electron spectrum, which would 
  quickly be eliminated by kinetic plasma processes. The spectral index
  increases linearly with $\log\eps$.
\item It allows a better comparison with acceleration models which naturally
  produce slightly curved electron spectra (like stochastic acceleration).
\end{enumerate}

Therefore, we fit the spectra to a photon model with an isothermal component at
lower energies and a log-parabolic component as given above at higher energies.
The background is taken into account in the following way: the pre-event and
post-event background spectra are measured and averaged (in some cases, particle
contamination prevented to obtain both of them, and only one was taken instead),
yielding a reference background spectrum. The reference background spectrum is
multiplied with a free parameter $\lambda$ and added to the model
spectrum. $\lambda$ is fitted together with the other model parameters. 

In large flares, sometimes an additional hard and weak $\gamma$-ray emission
from electrons is observed above 200 keV \citep{2008ApJ...678L..63K}. In this
paper we do not report on the properties of this emission since it is not
related to the hardening of the spectrum at lower energies. Does this component
negatively affects our fittings?  Examination of the fitted spectra reveal that
the fittings account for this weak, hard emission by an increase of the
parameter describing the strength of the background ($\lambda$) by a factor of
about 1 to 2. The fittings are of good quality because the background is also
weak and hard. This erroneous increase in the strength of the background has no
effect at lower energies, since there the hard X-ray flux is stronger by orders
of magnitude.

\section{Pivot point and parabolic fitting}
\label{app:b1}

A linear dependence of $\gamma$ vs. $\log I_{\eps_0}$ with negative slope can be
interpreted geometrically as a fixed intersection point of the various power-law
spectra at different times. The intersection in the spectral plane ($I$
vs. $\eps$) is called the \emph{pivot point}, located below the reference energy
$\eps_0$ \citep{grigis05}. Similarly, a linear dependence with positive slope
can be interpreted by a pivot point at energy larger than $\eps_0$. If fitting a
log-parabolic spectrum, it is the tangents to the spectrum at $\eps_0$ in
log-log space that are intersecting, rather than the curves themselves.

The description in terms of a pivot point has the advantage that it does not
depend on the choice of the reference energy $\eps_0$. On the other hand, the
pivot point energy jumps from $0$ to $+\infty$ when the slope in $\gamma$-$\log
I_{\eps_0}$ goes from negative to positive.  The relation between the pivot
point coordinates $\eps_*,I_*$ and the line parameters $a,b$ in
Eq. \ref{eq:lintrendglogf} are given by
\begin{equation}
\label{eq:pivslope}
\qquad a=\frac{1}{\log \left(\eps_*/\eps_0\right)}\qquad
\qquad b=\frac{-\log{I_*}}{\log \left(\eps_*/\eps_0\right)}\,.
\end{equation}

Since the spectra are curved in log-log space, the local spectral
index~$\gamma^\prime$, defined as the logarithmic derivative of the spectrum
\begin{equation}
\gamma^\prime(\eps)=\diff{\log I(\eps)}{\log \eps}=\gamma - 2\eta\log(\eps/\eps_0)\,,
\end{equation}
is energy dependent. The spectral parameter $\gamma$ is equal to the local
spectral index $\gamma^\prime$ at the reference energy $\eps_0=50$~keV. The
presence of a strong correlation of the time series of $\gamma$ and $\log
I_{50}$ does not necessarily imply a strong correlation of the time series of
$\gamma^\prime(\eps)$ and $\log I(\eps)$ at energies $\eps\ne \eps_0$.  In fact,
if the spectra have a common pivot point at $\eps_*<\eps_0$, the correlation of
$\gamma^\prime(\eps)$ and $\log I(\eps)$ weakens near the pivot point, and turns
into an anticorrelation for $\eps<\eps_*$. On the other hand, if a pivot point
exist at $\eps_*>\eps_0$, then the anticorrelation between $\gamma^\prime(\eps)$
and $\log I(\eps)$ for $\eps\simeq\eps_0<\eps_*$ transforms itself into a
correlation  at $\eps > \eps_*$.

An examination of the data for the events studied here shows that in the time
intervals when the correlation coefficient between $\log I_{50}$ and $\gamma$ is
near 1, the correlation coefficient between $\gamma^\prime(\eps)$ and $\log
I(\eps)$ is approximately constant at energies higher than 30-40 keV However, it 
shows a strong decrease to values close to -1 at lower energies such that the
transition takes place around 5-20 keV, near the pivot point position.  On the
other hand, when the correlation coefficient is near -1 at 50 keV we observe
that the correlation coefficients rises toward 1 at higher energies, as these
events tend to have a pivot point at energies larger than 50 keV.

Therefore, the energy dependence of the correlation in the data follows a
similar pattern as the one expected for non-curved spectra. This is
because the observed spectra are not strongly curved: $\eta<0.25\gamma$ in
all the fitted spectra, and $\eta<0.15\gamma$ in 85\% of all fitted spectra.

\section{Comparison between model and observation}
\label{app:b}

Comparison of non-thermal hard X-ray spectral observations and theoretical
models can be performed in different ways. In our case, we have a really simple
model and a time-dependent situation. The goal of the comparison is not the
perfect reproduction of every single observed spectrum, but rather a coherent
description of the time evolution which should be compatible with the observed
data. This is implemented by the additional constraint in the model (as given by
Eq. \ref{eqn:modelsp}) of keeping a constant electron spectral index $\delta$
above the threshold energy $E_\mathrm{T}$. So our task consists of two steps:
first, for a given value of $\delta$, we have to find the set of
$E_\mathrm{T}(t)$ and $F_{E_0}(t)$ that best reproduces the data and second, we
have to choose the best value of $\delta$ (for instance by running the first
step for many different values of $\delta$ and pick the best one).

For the comparison of the model with the data, we generate model photon spectra
emitted by the model electron spectra by computing the thick-target
Bremsstrahlung emission assuming collisional energy losses and using the full
relativistic Bethe-Heitler cross section \citep{BH1934} with the \cite{elwert39}
correction factor.

For the first step, the obvious strategy involves forward fitting the electron
model to the observed data. However, this is time consuming because it needs to
be repeated for all the spectra and many different values of $\delta$. On the
other hand, we have at our disposal the photon fitting parameters $I_{\eps_0}$,
$\gamma$ and $\eta$ (Eq. \ref{eq:spmodel}), which are good descriptions of the
observed photon spectrum as long as the reduced $\chi^2$ is around one. In this
case, there is no need of additional fitting in count space: we just fit exactly
the same log-parabolic model to the model photon spectra (in the energy range
where the non-thermal component is seen above the thermal component and the
background in the observations). This is faster, and delivers photon model
parameters $I^{\mathrm{MOD}}_{\eps_0}$, $\gamma^{\mathrm{MOD}}$ and
$\eta^{\mathrm{MOD}}$ as a function of the electron model parameter $F_{E_0}$,
$\delta$ and $E_\mathrm{T}$.  The comparison can then be performed in the
$\gamma$ vs. $I_{\eps_0}$ plot by a least square argument in two steps.

In the first step we held $\delta$ fixed and increase $E_\mathrm{T}$ to
  generate a curve in the $\gamma$ vs. $I_{\eps_0}$ plot for each value of
$F_{E_0}$. The normalization $F_{E_0}$ is then chosen such that it minimizes the
square differences between $I_{\eps_0}$ and $I^{\mathrm{MOD}}_{\eps_0}$. The
results from running step one repeatedly are a set of paths (one for each
different value of $\delta$) in the $\gamma$ vs. $I_{\eps_0}$ plot, where
the variable $E_\mathrm{T}$ changes along the path.

In the second step, we just select among all paths in the $\gamma$
vs. $I_{\eps_0}$ space the one with the least square distances from all the
observed points. This yields the best $\delta$ and $F_{E_0}$ and a range of
variation of $E_\mathrm{T}$.

%

%

\begin{thebibliography}{}
%

 
\bibitem[Bai(1986)]{bai1986} Bai, T.\ 1986, \apj, 308, 912 

\bibitem[Bai \& Ramaty(1979)]{bai1979} Bai, T., \& Ramaty, R.\ 1979, \apj, 227, 1072

\bibitem[Battaglia \& Benz (2007)]{2007A&A...466..713B}
 Battaglia, M., \& Benz, A.~O.\ 2007, \aap, 466, 713

\bibitem[Battaglia \& Benz (2006)]{battaglia06}
 Battaglia, M \& Benz, A.~O. 2006, \aap, 456, 751

\bibitem[Battaglia \etal (2005)]{battaglia05}
 Battaglia, M., Grigis, P.~C., \& Benz, A.~O. 2005, \aap, 439, 737 

\bibitem[Bethe \& Heitler (1934)]{BH1934}
 Bethe, H., \& Heitler, W. 1934, Royal Society of London Proceedings Series~A,
 146, 83

\bibitem[Cliver \etal (1986)]{cliver86}
 Cliver, E.~W., Dennis, B.~R., Kiplinger, A.~L., et al. 1986, \apj, 305, 920 

\bibitem[Dulk \etal (1992)]{dulk92}
 Dulk, G.~A., Kiplinger, A.~L., \& Winglee, R.~M.\ 1992, \apj, 389, 756

\bibitem[Frost \& Dennis (1971)]{frost71}
 Frost, K.~J., \& Dennis, B.~R.\ 1971, \apj, 165, 655

\bibitem[Emslie \etal (2004)]{emslie04}
 Emslie, A.~G., Miller, J.~A., \& Brown, J.~C.\ 2004, \apjl, 602, L69

\bibitem[Elwert (1939)]{elwert39}
 Elwert, G. 1939, Ann. Phys. 34, 178

\bibitem[Grigis \& Benz (2004)]{grigis04}
 Grigis, P.~C. \& Benz, A.~O. 2004, A\&A, 426, 1093

\bibitem[Grigis \& Benz (2005a)]{grigis05}
 Grigis, P.~C., \& Benz, A.~O.\ 2005a, \aap, 434, 1173 

\bibitem[Grigis \& Benz (2005b)]{2005ApJ...625L.143G}
 Grigis, P.~C., \& Benz, A.~O.\ 2005b, \apjl, 625, L143

\bibitem[Grigis \& Benz (2006)]{grigis06}
 Grigis, P.~C., \& Benz, A.~O.\ 2006, \aap, 458, 641

\bibitem[Hudson \etal (1982)]{hudson82}
 Hudson, H.~S., Lin, R.~P., \& Stewart, R.~T.\ 1982, \solphys, 75, 245

\bibitem[Kahler (1984)]{kahler84}
 Kahler, S.~W.\ 1984, \solphys, 90, 133

\bibitem[Kiplinger (1995)]{kiplinger95}
 Kiplinger, A.~L.\ 1995, \apj, 453, 973 

\bibitem[Krucker et al.(2008)]{2008ApJ...678L..63K}
 Krucker, S., Hurford, G.~J., MacKinnon, A.~L., Shih, A.~Y., \& Lin, R.~P.\ 2008, \apjl, 678, L63 


\bibitem[Lin \etal (2002)]{lin02}
 Lin, R.~P., Dennis, B.~R., Hurford, G.~J., et al. 2002, Sol. Phys., 210, 3

\bibitem[Massaro \etal (2000)]{massaro00}
 Massaro, E., Cusumano, G., Litterio, M., \& Mineo, T.\ 2000, \aap, 361, 695 

\bibitem[Ohki \etal (1983)]{ohki83}
 Ohki, K., Takakura, T., Tsuneta, S., \& Nitta, N.\ 1983, \solphys, 86, 301

\bibitem[Parks \& Winckler (1969)]{parks69}
 Parks, G.~K., \& Winckler, J.~R.\ 1969, \apjl, 155, L117

\bibitem[Saldanha \etal (2008)]{2008ApJ...673.1169S}
 Saldanha, R., Krucker, S., \& Lin, R.~P.\ 2008, \apj, 673, 1169 

\bibitem[Piana et al.(2003)]{2003ApJ...595L.127P}
  Piana, M., Massone, A.~M., Kontar, E.~P., Emslie, A.~G., Brown, J.~C., \&
  Schwartz, R.~A.\ 2003, \apjl, 595, L127

\bibitem[Simnett (2006)]{2006A&A...445..715S}
 Simnett, G.~M.\ 2006, \aap, 445, 715

\bibitem[Takakura \etal (1984)]{takakura84}
 Takakura, T., Sakurai, 
 T., Ohki, K., Wang, J.~L., Zhao, R.~Y., Xuan, J.~Y., \& Li, S.~C.\ 1984, 
\solphys, 94, 359 

\bibitem[Wild \etal (1963)]{wild63}
 Wild, J.~P., Smerd, S.~F., \& Weiss, A.~A.\ 1963, \araa, 1, 291

\bibitem[Yashiro \etal (2004)]{2004JGRA..10907105Y}
Yashiro, S., Gopalswamy, N., Michalek, G., St.~Cyr, O.~C., Plunkett, S.~P.,
Rich, N.~B., \& Howard, R.~A.\ 2004, Journal of Geophysical Research (Space
Physics), 109, 7105


\end{thebibliography}
\end{document}